\documentclass[aps,pre,twocolumn,showpacs]{revtex4}
\usepackage{epsfig}
\usepackage{times}
\usepackage{amsmath}

\begin{document}

\title{Impact of aging on the evolution of cooperation in the spatial prisoner's dilemma game}

\author{Attila Szolnoki$^1$, Matja{\v z} Perc$^2$, Gy{\"o}rgy Szab{\'o}$^1$ and Hans-Ulrich Stark$^{3,4}$}
\affiliation
{$^1$Research Institute for Technical Physics and Materials Science, P.O. Box 49, H-1525 Budapest, Hungary\\$^2$Department of Physics, Faculty of Natural Sciences and Mathematics, University of Maribor, Koro{\v s}ka cesta 160, SI-2000 Maribor, Slovenia\\$^3$swissQuant Group AG, Universit{\"a}tstrasse 9, CH-8006 Zurich, Switzerland\\$^4$Chair of Systems Design, ETH Zurich, Kreuzplatz 5, CH-8032 Zurich, Switzerland}

\begin{abstract}
Aging is always present, tailoring our interactions with others and postulating a finite lifespan during which we are able to exercise them. We consider the prisoner's dilemma game on a square lattice, and examine how quenched age distributions and different aging protocols influence the evolution of cooperation when taking the life experience and knowledge accumulation into account as time passes. In agreement with previous studies, we find that a quenched assignment of age to players, introducing heterogeneity to the game, substantially promotes cooperative behavior. Introduction of aging and subsequent death as a coevolutionary process may act detrimental on cooperation but enhances it efficiently if the offspring of individuals that have successfully passed their strategy is considered newborn. We study resulting age distributions of players, and show that the heterogeneity is vital yet insufficient for explaining the observed differences in cooperator abundance on the spatial grid. The unexpected increment of cooperation levels can be explained by a dynamical effect that has a highly selective impact on the propagation of cooperator and defector states.
\end{abstract}

\pacs{87.23.-n, 02.50.Le, 87.23.Ge}

\maketitle

\section{Introduction}

If individual interests are in dissonance with the collective wellbeing a social dilemma is in place \cite{macy02}. Inherently driven by the fundamental principles of Darwinian selection, social dilemmas constitute an intriguing puzzle that can be faced across the whole of social and natural sciences \cite{axelrod84}. Cooperative behavior, implying working for the common good of a society against the innate selfish drive that is routed in each individual, promises a departure from the impending social decline. The subtleties of cooperation within groups of selfish individuals are most frequently investigated within the evolutionary game theory \cite{weib95, gintis00, nowak06, szabo07}. Particularly the prisoner's dilemma game seems best suited to address the issue. In it's original form the game consists of two players who have to decide simultaneously whether they want to cooperate or defect. The dilemma is given by the fact that although mutual cooperation yields the highest collective payoff, individual defectors will do better if the opponent decides to cooperate. Since selfish players are aware of this fact they both decide to defect, thus constituting overall defection as the Nash equilibrium of the prisoner's dilemma game \cite{hofbauer98}. Prominently, spatial structure may foster the formation of cooperative clusters on the grid \cite{nowak92, szabo98, hauert04} and thus, depending on the temptation to defect, sustain some fraction of the population in a cooperative state.

The subject of spatial evolutionary games was extended by introducing complex networks as the underlying interaction topology of players \cite{abramson01, holme03, wu05, wang_pre08, szabo05, wang06, fu_pre08, poncela07, wang07, rong07, ohtsukiprl07, chen08, assenza08, xiaojie_pre09, fu_pre09}, whereby outstanding is the realization that highly heterogeneous scale-free networks provide a near optimal environment for a flourishing cooperative state \cite{santos05}. The important role of heterogeneities within evolutionary games on complex networks has been additionally amplified by studies considering participation costs or the usage of normalized or effective payoffs \cite{santosjb06, tomassini07, masuda07, szolnoki08}. Notably, heterogeneities can also be introduced via differences in the influence and strategy transfer capability of players \cite{kim02, wu06, szolnoki07, guan07} or social diversity \cite{perc08, santos08, chen_pre09}, whereby the impact on the evolution of cooperation was found to be similarly beneficial as in the context of studies considering evolutionary games on complex network. Recently, however, the focus of research activity has been shifting towards coevolutionary mechanisms that are able to generate the heterogeneities necessary for the promotion of cooperation spontaneously. In particular, the aim is not to introduce the heterogeneities artificially through complex interaction networks, differences in the strategy transfer capability or social diversity, but to let them evolve alongside the main evolutionary process of strategy adoption with coevolutionary rules that, on their own, do not violate the rank of participating strategies. For example, in \cite{szolnokinjp08} the strategy transfer capability (or teaching activity) was considered as an evolving property of players, and it has been shown that simple coevolutionary rules may lead to highly heterogeneous distributions from an initially non-preferential setup, in turn promoting cooperation in social dilemmas. Moreover, it has recently been shown that highly heterogeneous interaction networks may evolve spontaneously from simple coevolutionary rules \cite{lipre07, poncelaplos08, szolnokiepl08}, and processes like prompt reaction to adverse ties \cite{segbroeckbmc08, segbroeckprl09} or reputation-based partner choice \cite{fupre08} have all been considered as coevolutionary rules that can promote cooperative behavior. Interestingly, mobility may also be considered as a coevolutionary process, and indeed, recent studies have shown \cite{vainsteinjtb07, helbing_pnas09, helbing_epjb09} that it may have a beneficial impact on the evolution of cooperation. Pioneering in view of introducing coevolutionary processes to evolutionary game theory have been works studying the impact of active or dynamical linking \cite{pachecojtb06, pachecoprl06}, as well as earlier studies considering random or intentional rewiring procedures \cite{zimmermann04, zimmermann05, percnjp06}.

Presently, we study the impact of artificially introduced heterogeneity, and heterogeneities arising from a non-preferential deterministic aging protocol, on the evolution of cooperation in the prisoner's dilemma game on a square lattice. Alternatively, the deterministic rule for aging can be considered as a coevolutionary process, during which age as a property of each player changes alongside the abundance of the two participating strategies. The two approaches have a joint root taking explicitly into account the age of participating players, thus offering a unique constellation enabling us to investigate differences between artificially and spontaneously introduced heterogeneities within the context of evolutionary game theory. Thereby, age as a property, and aging as a process, both seem to be very natural ingredients that relevantly enrich the main evolutionary process of strategy adoption on the spatial grid. Here it is worth mentioning that the positive effect of age (and memory-dependent transition rates) on the consensus formation within the voter model has been described recently in \cite{stark08a, stark08b}, and
through the spatial effect this mechanism can help to maintain cooperative behavior.

Since age is often associated with knowledge and wisdom an individual is able to accumulate over the years, we introduce it to the studied prisoner's dilemma game through a simple tunable function that maps age to teaching activity (henceforth strategy transfer capability) of the corresponding player. According to logical reasoning, we consider older players to be more knowledgable than younger individuals, and the former are thus also characterized with a higher strategy transfer capability and related reproduction probability \cite{szolnoki07}. In case age is assigned artificially (and randomly) to each player following a uniform distribution and does not evolve in time (quenched), we find that cooperation promotion depends significantly on the level of heterogeneity that is introduced through the function that maps age to strategy transfer capability, similarly as reported recently in \cite{perc08}. However, in case of the most successful aging protocol, when strategy adoption is accompanied with the emergence of a newborn, the resulting strategy pass capability distribution does not differ relevantly enough from the previously assumed distribution to explain the enhanced success of the mentioned protocol. As we will show, the significant improvement of cooperation promotion must be attributed to the details of a microscopic mechanism that promotes the propagation of cooperator and defector strategies in a highly selective manner. Thus, we report that, not only may simple coevolutionary rules spontaneously generate highly heterogeneous states that, on their own, substantially promote cooperation, but may also affect the strategy adoption process on a player-to-player level, which provides an additional and unsuspected lift to the cooperative trait.

The remainder of this paper is organized as follows. In the next section we describe the evolutionary prisoner's dilemma game and the aging protocols. Section 3 is devoted to the presentation of results, whereas lastly we summarize and discuss their implications.

\section{Game definitions and aging}

We consider an evolutionary prisoner's dilemma game that is characterized with the temptation to defect $T = b$ (the highest payoff received by a defector if playing against a cooperator), reward for mutual cooperation $R = 1$, and the punishment for mutual defection $P$ as well as the suckers payoff $S$ (the lowest payoff received by a cooperator if playing against a defector) equaling $0$. Thereby $1 < b \leq 2$ ensures a proper payoff ranking and preserves the essential dilemma between individual profits and welfare of the population for repeated games \cite{nowak92}. This choice is motivated with the aim of studying a simple and frequently adapted model, but we note that our findings are robust and can be observed in the full two-parameter prisoner's dilemma game as well \cite{santospnas06}.

Throughout this work each player $x$ on the regular $L \times L$ square lattice is connected to its four nearest neighbors and initially designated either as a cooperator ($s_x=C$) or defector ($D$) with equal probability, and the game is iterated forward in accordance with the Monte Carlo simulation procedure comprising the following elementary steps. First, a randomly selected player $x$ acquires its payoff $p_x$ by playing the game with its nearest neighbors. Next, one randomly chosen neighbor, denoted by $y$, also acquires its payoff $p_y$ by playing the game with its four neighbors. Lastly, player $x$ tries to enforce its strategy $s_x$ on player $y$ in accordance with the probability
\begin{equation}
W(s_x \rightarrow s_y)=w_x \frac{1}{1+\exp[(p_y-p_x)/K]},
\label{eq:prob}
\end{equation}
where $K$ denotes the amplitude of noise \cite{szabo98} or its inverse ($1/K$) the so-called intensity of selection \cite{traulsen07, altrock09}, and $w_x$ characterizes the strategy transfer capability of player $x$ \cite{szolnoki07}. One full Monte Carlo step involves all players having a chance to pass their strategies to their neighbors once on average. To introduce the previously mentioned difference between young and old players, the strategy transfer capability $w_x$ is presently related to the integer age $e_x=0, 1, \ldots, e_{max}$ in accordance with the function $w_x=(e_x/e_{max})^\alpha$, where $e_{max}=99$, denoting the maximal possible age of a player, serves the bounding of $w_x$ to the unit interval, and $\alpha$ determines the level of heterogeneity in the $e_x \rightarrow w_x$ mapping. Evidently, $\alpha=0$ corresponds to the classical (homogeneous) spatial model with $w_x=1$ characterizing all players, $\alpha=1$ ensures that $w_x$ and $e_x$ have the same distribution, whereas values of $\alpha \geq 2$ impose a power law distribution of strategy transfer capability too. We note, however, that at larger values of $\alpha$ the distribution of $w_x$ becomes so heterogeneous that the majority of players is unable to pass their strategy. This may cause frozen states or extremely long relaxation times. To avoid either of the two, we choose $\alpha=2$ as the highest value in this study. It is also worth noting that, according to the mapping, larger values of $e_{max}$ may result in more heterogeneous states than lower $e_{max}$ (by a given $\alpha$). Thus, in accordance with previous findings \cite{perc08}, the borders of $b$ where cooperators survives may shift higher. However, since other essentials of below presented results are thereby not affected, and in order to focus on the new features of the model, we keep $e_{max}=99$ constant throughout this work.

Here we separately consider the case where initially all $e_x$ are selected randomly from a uniform distribution within the interval $[0,e_{max}]$ and do not evolve in time, whereby $\alpha$ determines the level of artificially introduced heterogeneity as described. Moreover, we also study the coevolutionary model incorporating aging, death and newborns. The coevolutionary aging protocol entails; starting from the same age distribution as supposed previously, the age of all players is increased by $1$ for each Monte Carlo step (MCS), furthermore, setting $e_x=0$ for all players $x$ whose age exceeded $e_{max}$ (effectively this means that a newborn follows the dead player). Importantly, within this coevolutionary model we have studied two options to handle the age of players that have just adopted a strategy from one of their neighbors. Either their age may be left unchanged (coevolutionary model A) or they can be considered as newborns (coevolutionary model B), \textit{i.e.} as soon as player $x$ adopts a strategy its age is set to $e_x=0$. Notably, models A and B can be interpreted rather differently. From a purely biological viewpoint the more successful player replaces the neighbor with its own offspring, who therefore initially has a limited strategy transfer capability, which corresponds to model B. On the other hand, especially in social systems, strategy adoptions may not necessarily involve death and newborns, but may indicate solely a change of heart, preference, or way of thinking, whereby this situation corresponds to model A. Nevertheless, newborns in a social context can be considered those that changed their strategy recently, and therefore have a low reputation initially. We will study models A and B separately, and show that the concept of solely strategy adoption and the seemingly similar reproduction with an offspring (entailing also the death of the previous player) may have very different impacts on cooperation within coevolutionary game theory models.

Results of Monte Carlo simulations presented below were obtained on populations comprising $100 \times 100$ to $800 \times 800$ individuals (additional details are provided in the figure captions), whereby the stationary fraction of cooperators $\rho_C$ was determined within $10^5$ to $10^6$ full MCS after sufficiently long transients were discarded. Moreover, since the coevolutionary aging process may yield highly heterogeneous distribution of $e_x$, which may be additionally amplified during the $e_x \rightarrow w_x$ mapping, final results were averaged over up to $20-300$ independent runs for each set of parameter values in order to assure suitable accuracy.

\section{Results}

\begin{figure}
\begin{center} \includegraphics[width = 8cm]{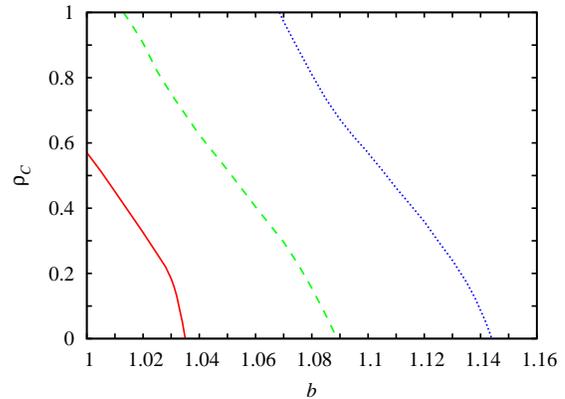}
\caption{\label{fig1}(color online) Promotion of cooperation due to the increasing heterogeneity in the $e_x \rightarrow w_x$ mapping via $\alpha$. Stationary fraction of cooperators $\rho_C$ is plotted in dependence on $b$ for $\alpha=0$ (solid red line), $\alpha=1$ (dashed green line) and $\alpha=2$ (dotted blue line). In all three cases $K=1$. The typical system size at this noise level was $L=400-800$.}
\end{center}
\end{figure}

\begin{figure}
\begin{center} \includegraphics[width = 8cm]{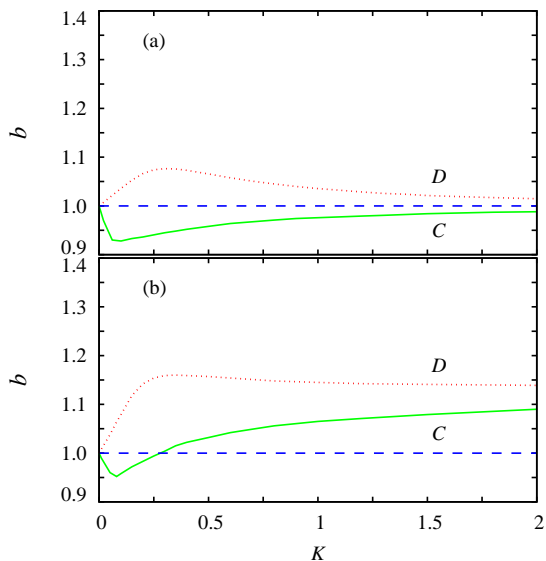}
\caption{\label{fig2}(color online) Full $b-K$ phase diagrams for the prisoner's dilemma game with quenched uniform distribution of $e_x$, obtained by setting $\alpha=0$ [panel (a)] and $\alpha=2$ [panel (b)] in the $e_x \rightarrow w_x$ mapping. Solid green and dotted red lines mark the borders of pure $C$ ($\rho_C=1$) and $D$ ($\rho_C=0$) phases, respectively, whereas in-between a mixed phase characterizes the distribution of strategies on the spatial grid. Dashed blue line at $b=1$ denotes the end of the prisoner's dilemma payoff parametrization. The phase transition points were determined using $L=400-800$.}
\end{center}
\end{figure}

We start by presenting results obtained with the quenched age model, where initially all $e_x$ are selected randomly from a uniform distribution and do not evolve in time. Figure~\ref{fig1} shows how $\rho_C$ varies in dependence on the temptation to defect $b$ for three different values of $\alpha$. It can be observed that cooperation is promoted more effectively as $\alpha$ increases, which is in agreement with the fact that $\alpha=0$ returns the classical spatial prisoner's dilemma game, $\alpha=1$ simply copies the uniform distribution of $e_x$ onto $w_x$, and $\alpha=2$ transform the uniform distribution of $e_x$ to power law distributed $w_x$ with a slope on a double logarithmic graph equalling $-0.5$. Since an increase of $\alpha$ thus directly implies an increase in the heterogeneity of players on the spatial grid, these results confirm the argumentation presented recently in \cite{perc08}. To further establish the fact that highly heterogeneous states promote cooperation, we present in Fig.~\ref{fig2} full $b-K$ phase diagrams for $\alpha=0$ [panel (a)] and $\alpha=2$ [panel (b)]. Evidently, the promotive impact prevails across the whole span of $K$, enhancing not just the mixed phase region (area between the dashed green and the solid red line), but also the extend of complete cooperator dominance (area below the dashed green line), which is fully absent in the $\alpha=0$ case provided the prisoner's dilemma payoff parametrization is considered ($b=1$ denoted by the dashed blue line).

\begin{figure}
\begin{center} \includegraphics[width = 8cm]{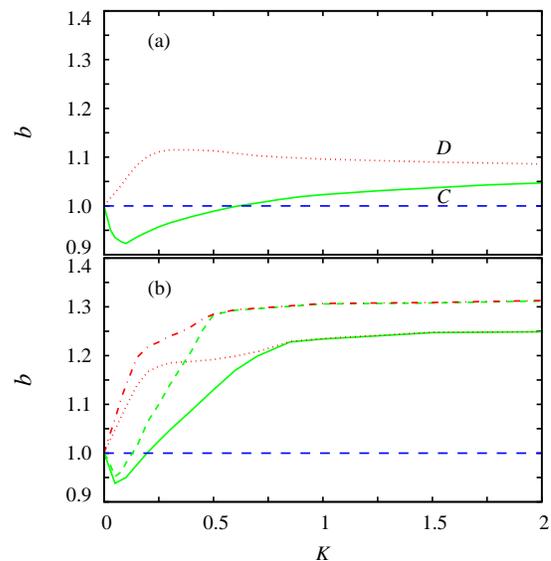}
\caption{\label{fig3}(color online) Full $b-K$ phase diagrams for the prisoner's dilemma game incorporating aging as a coevolutionary process, obtained by setting $\alpha=2$ and considering players who have adopted a strategy as newborns [model B - their age is set to $e_x=0$; panel (b)] or not [model A - $e_x$ is unaffected by the strategy adoption; panel (a)]. Solid green and dotted red lines mark the borders of pure $C$ ($\rho_C=1$) and $D$ ($\rho_C=0$) phases, respectively, whereas the intermediate region characterizes the coexistence of $C$ and $D$ strategies on the spatial grid. Additionally, panel (b) features the phase diagram (depicted by dashed green and dash-dotted red lines) obtained if the time scales of aging and strategy adoption are separated (aging is $90 \%$ slower), whereby the colors of the lines have the same meaning as before [note that $C$ and $D$ symbols are not plotted in panel (b) to avoid ambiguity]. Dashed blue line has the same meaning as in Fig.~\ref{fig2}.}
\end{center}
\end{figure}

Figure~\ref{fig3} shows Monte Carlo results on the same $b-K$ phase diagrams as presented in Fig.~\ref{fig2} for the above-mentioned coevolutionary aging protocols. More precisely, panel (a) shows results obtained via the coevolutionary model A where the age of players imitating a strategy remains unchanged, whereas panel (b) depicts results of model B where players adopting a new strategy are considered as newborns (their age becomes $e_x=0$ as soon as they adopt the new strategy). Although in both cases the mapping from $e_x$ to $w_x$ is realized by using $\alpha=2$, and the two coevolutionary models seem to differ only minutely, the models A and B offer very different levels of support for cooperative behavior. While the coevolutionary model A performs worse than the quenched age model when using the same $\alpha=2$ [for the sake of comparison we used the same horizontal and vertical scales in Fig.~\ref{fig2}(b)], the coevolutionary model B surpasses its cooperation promotion abilities markedly. Notice that the level of cooperation can be enhanced further if the time scales of aging and strategy adoption are separated \cite{sanchez06}, for example by letting only $10 \%$ of randomly chosen players (instead of all) increase their age after each MCS. The phase diagram depicted by dashed green and dash-dotted red lines in Fig.~\ref{fig3}(b) highlights an example of time scale separation impact on the evolution of cooperation within model B. However, due to extremely sharp phase transitions (note that for all $K>0.3$ the mixed $C+D$ phase is virtually absent), and long relaxation times associated with coevolutionary models that have separated time scales, we restrain our analysis to the originally proposed aging protocols as described in Section 2.

\begin{figure}
\begin{center} \includegraphics[width = 8cm]{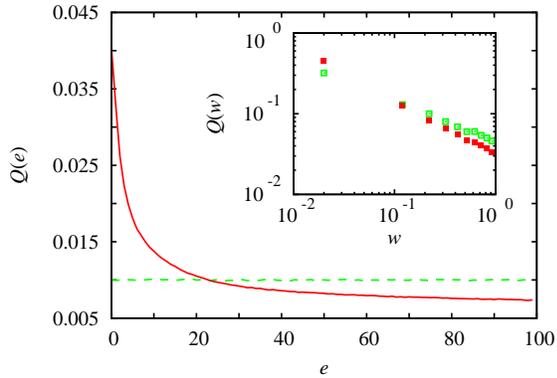}
\caption{\label{fig4}(color online) Final distributions of age $Q(e)$ for the prisoner's dilemma game incorporating aging as a coevolutionary process, obtained by considering players who have adopted a strategy as newborns (model B; solid red line) or not (model A; dashed green line; note that the quenched age model is characterized by an identical uniform distribution). The inset features the distributions of strategy transfer capability $Q(w)$ in the corresponding colors for $\alpha=2$. Note that both axes in the inset have a logarithmic scale, and that thus the depicted linear dependencies correspond to power law distributed values of $w$. Importantly, the slope of a liner function fitted to green symbols has slope $-0.5$, whereas the slope for the red symbols is only marginally higher, equalling $-0.7$. The employed system size was $L=200$.}
\end{center}
\end{figure}

In order to explain the differences in cooperation promotion observed in Fig.~\ref{fig2}(b) and Figs.~\ref{fig3}(a) and (b) (note that in all there cases we used $\alpha=2$ in the $e_x \rightarrow w_x$ mapping), we examine the probability distributions of resulting age ($Q(e)$) and corresponding strategy transfer capabilities ($Q(w)$) due to the two proposed coevolutionary aging protocols. Figure~\ref{fig4} depicts the obtained results. It is obvious that the coevolutionary model A has the same uniform distribution of $e_x$, and the corresponding power law distribution of $w_x$ for $\alpha=2$, as was also used in the quenched age model in Fig.~\ref{fig2}(b). On the other hand, the coevolutionary model B results in a substantially more heterogeneous age distribution, which however, results only in a modestly steeper power law distribution of $w_x$ (see the caption and inset of Fig.~\ref{fig4} for details). Thus, from the depicted distributions of strategy transfer capability alone, one would anticipate that the quenched age model and the coevolutionary model A would warrant equal promotion of cooperation, whereas only the coevolutionary model B would perform marginally better. However, neither of the two statements are very accurate, since in fact model A promotes cooperation worse than the quenched age model, while model B outperforms both by a significant margin [compare the corresponding phase diagrams depicted in Fig.~\ref{fig2}(b) and Figs.~\ref{fig3}(a) and (b)]. From this we conclude that, although heterogeneity arguably plays a crucial role in promoting cooperation \cite{santos05, perc08} and can evolve spontaneously from strikingly simple strategy independent aging protocols, coevolutionary rules may trigger additional mechanisms that work either in favor or against cooperative behavior yet cannot be detected effectively by global statistical measures.

\begin{figure}
\begin{center} \includegraphics[width = 8cm]{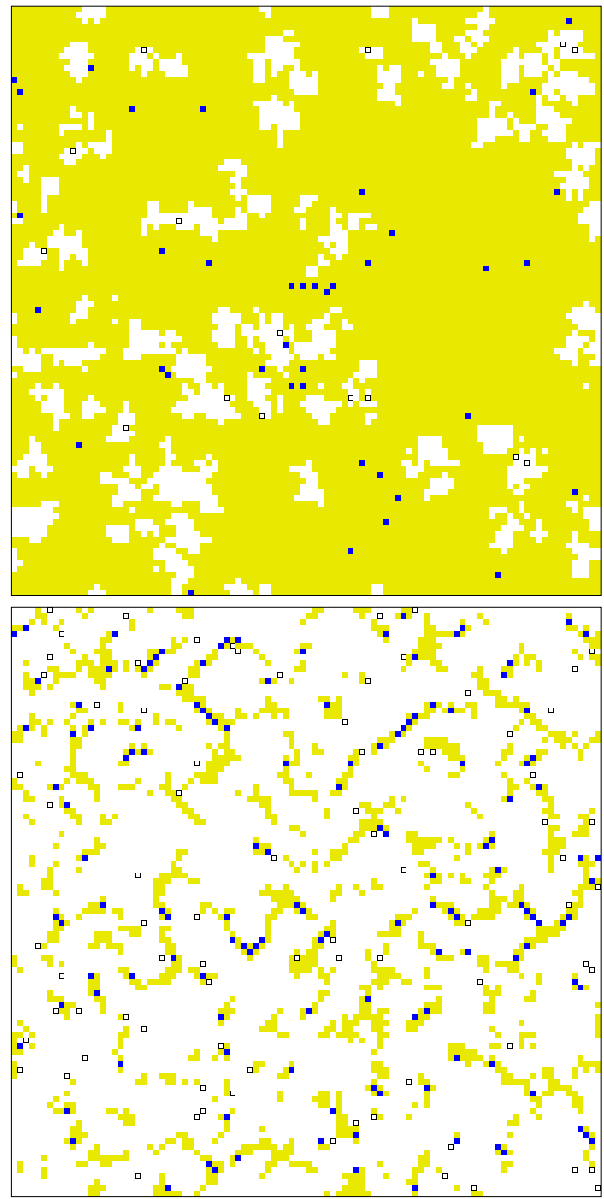}
\caption{\label{fig5}(color online) Snapshots of typical distributions of players on a $100 \times 100$ spatial grid, obtained by considering players who have adopted a strategy as newborns (model B; bottom panel) or not (model A; top panel). Blue closed and black open boxes depict influential players (see text for details) in defector and cooperator states, respectively; while yellow and white are all the other non-influential defectors and cooperators, respectively. Employed parameter values for both snapshots are: $K=0.1$, $b=1.05$ and $\alpha=2$.}
\end{center}
\end{figure}

Indeed, by closely examining the impact of the coevolutionary aging protocols on the microscopic player-to-player level, thus far hidden secondary mechanisms become visible that may either hinder (as by model A) or promote (as by model B) the evolution of cooperation beyond the levels indicated by heterogeneity alone. For model A the moderate decrease of cooperation levels, if compared to the quenched age model, can be explained by the fact that cooperative domains, created around players with high strategy transfer capabilities, cannot prevail long. Namely, the central cooperator who built up the cooperative domain eventually dies, and the arriving newborn with an accordingly low strategy transfer capability simply cannot maintain this domain further, thus giving defectors an opportunity to win it over. However, by the B model the situation changes significantly because the central players are always surrounded by newborns. Thereby it is important to note that whenever an old defector, with a high strategy transfer capability, is imitated by one of the neighbors, further spreading of defection is blocked because the newborn defector has no chance to pass strategy $D$ further. At that time a neighboring cooperator with high age can strike back and conquer the site of the newborn defector. As a result the whole procedure starts again, which ultimately results in a practically blocked (more precisely an oscillating) front between $C$ and $D$ regions. Crucially, a similar blocking mechanism is not present around old (and thus influential) cooperators because there cooperator-cooperator links help newborn cooperators to achieve higher age, in turn supporting the overall maintenance of cooperative behavior. This phenomenon is nicely illustrated by the comparative snapshots in Fig.~\ref{fig5}, where the so-called influential players with $C$ and $D$ strategies are indicated by blue closed and black open boxes, respectively. Irrespective of the strategy, a player is designated as being influential if its age exceeds that of any of its neighbors by at least $e_{max}/2$ (notably, qualitatively similar snapshots can be obtained by choosing different thresholds as well). It can  be observed clearly that influential defectors (blue closed boxes) in the bottom panel are surrounded by narrow yellow stripes (non-influential defectors), thus reflecting the above-described blocked propagation of defector states. On the other hand, large homogeneous white regions indicate that players within cooperative domains can age together without extinction, and therefore the age difference between neighbors does not grow permanently. Conversely, the influential defectors in the top panel of Fig.~\ref{fig5} (coevolutionary model A) don't experience propagation restrictions, and accordingly, can easily spread their strategy across the majority of the spatial grid. Notice furthermore that the overall number of influential player is strikingly larger in the B model than in the A model. This can be understood by the frequent strategy changes occurring in the neighborhoods of defectors.

The above snapshots visualize that aging as an unbiased coevolutionary process entailing newborn offspring introduces a highly selective mechanism favoring the propagation of cooperation at the expense of defection. In other words, the mentioned microscopic mechanism enhances the effect of inhomogeneities in the strategy transfer capabilities and provides a substantial boost for the cooperative behavior.

Finally, we mention that non-monotonous mappings between $e_x$ and $w_x$ can also be applied successfully within the proposed model. From the sociological or biological viewpoint the model can thus be extended to account for the fact that the oldest individuals may not necessarily be the most influential. Keeping the plausible condition that very young players have no or very little influence, the reduced influence of very old players is viable as well. For this purpose, we introduced a Gaussian distribution of $w_x$ having the maximum at an intermediate value of $e_x=80$ with variance $\sigma^2=350$. If the condition that suppresses the influence of newborns is fulfilled (their age becomes zero), the observed mechanism still works exactly as described for the power law distribution. Nevertheless, as could be anticipated, differences appear in dependence on the location of the peak as well as the width, yet they are limited to modest shifts of critical values of $b$. We also note that, if the distribution is narrowly centered (\textit{e.g.} $\sigma^2=25$) on a given $e_x$ (thus, the majority is unable to pass strategy), the relaxation times become extremely long. For example, even at small system sizes ($L=100$) $MCS=2 \cdotp 10^7$ may be a too short relaxation time, and thus one is faced with the same difficulties as appear if large $\alpha$ are used, as mentioned earlier. We thus warn from extensions in this direction and advise to handle resulting models with care an patience.

\section{Summary and discussion}

In sum, we have studied the impact of heterogeneities, motivated by different quenched or evolving ages of players, on the evolution of cooperation within the prisoner's dilemma game on a square lattice. We have established that by quenched age distributions the enhanced heterogeneities are advantageous for cooperative behavior. Moreover, we have shown that simple strategy independent coevolutionary aging protocols, supplemented by an appropriate function that maps age to teaching activity, may lead to highly heterogeneous states from an initially fully non-preferential setup, and thus significantly elevate the density of cooperators on the spatial grid. Importantly, however, we have also demonstrated that by certain coevolutionary rules heterogeneity alone may be an insufficient indicator of the actual cooperation promoting potential of the environment. In particular, by considering players who have adopted a strategy as newborns rather than individuals just changing their strategy, a new and powerful mechanism for promotion of cooperation is triggered that acts solely on the microscopic player-to-player basis, and is thus virtually non-detectable by statistical methods assessing the heterogeneity of the system. The mechanism relies on a highly selective promotion of cooperator-cooperator and defector-defector pairs, which hinders influential defectors to spread their strategy effectively across the spatial grid. As a consequence of the dynamical origin of the observed cooperation-promoting mechanism, it is expected that it will work in other cases too, for example when the interaction graph is characterized by a different topology.

Our findings are not restricted to the specific strategy transfer capability profile we generally used throughout this study, but remain valid for other distributions as well. Specifically, the distributions need not be monotonous, but may be bell-shaped with arbitrary skewness as well. The only condition that needs to be fulfilled is to keep the influence of newborn and young players at a low level. Evidently, the revealed strategy propagation selection mechanism can work more effectively if the interval of age that maps to low strategy transfer capabilities is longer.

Presented results outline new ways of promoting cooperative behavior via simple and natural coevolutionary rules that are able to exploit defector's weaknesses at the very beginnings of their emergence. From the real-life point of view, our findings support the notion that differences in age alone may have a beneficial impact on the evolution of cooperation, which might be particularly important for animal and ancient human societies, where the options for heterogeneities to emerge through workings made possible by advancements in technology have been or still are limited.

\begin{acknowledgments}
This work was supported by the Hungarian National Research Fund (grant K-73449) and the Bolyai Research Scholarship. MP acknowledges financing by the Slovenian Research Agency (grant Z1-2032-2547), whereas HUS acknowledges financial support from SBF Switzerland through the research project C05.0148 (Physics of Risk). The joint collaboration of AS, GS and MP is also supported by the Slovene-Hungarian bilateral project BI-HU/09-10-001.
\end{acknowledgments}

\end{document}